\documentclass[twocolumn,a4paper]{article}
\usepackage[colorlinks, urlcolor=blue,
	pdffitwindow, bookmarksopen,
	filecolor=blue,
	bookmarksopenlevel=0,
	pdfpagelayout=SinglePage, dvipdfm]{hyperref}
\usepackage{ncsp23} 
\usepackage{color}
\usepackage[dvipdfmx]{graphicx}
\usepackage{hyperref}

\begin{document}
\title{Forex Trading Strategy That Might Be Executed Due to the Popularity of Gotobi Anomaly}

\name{
\begin{tabular}{c}
Hiroki Bessho$^1$, Takanari Sugimoto$^2$, Tomoya Suzuki$^1$
\end{tabular}
}

\address{
\begin{tabular}{c}
$^1$
Major in Mechanical Systems Engineering, Graduate School of Science and Engineering, Ibaraki University\\
4-12-1 Nakanarusawa-cho, Hitachi-shi, Ibaraki, 316-8511 Japan. Phone: +81-294-38-5195\\
E-mail: \{21nm480s, tomoya.suzuki.lab\}@vc.ibaraki.ac.jp\\
$^2$
Derivatives Business Department, Gaika ex byGMO, Inc.
E-mail: takanari\_sugimoto@gaikaex.jp
\end{tabular}
}

\maketitle
\section*{Abstract}
Our previous research has confirmed that the USD/JPY rate tends to rise toward 9:55 every morning in the Gotobi days,
which are divisible by five.
This is called the Gotobi anomaly.
In the present study, we verify the possible trading strategy and its validity
under the condition that investors recognize the existence of the anomaly.
Moreover, we illustrate the possibility that the wealth of Japanese companies might leak to FX traders due to the arbitrage opportunity
if Japanese companies blindly keep making payments in the Gotobi days as a business custom.
%

\section{Introduction}
According to the efficient market hypothesis\cite{Fama} and the random walk theory,
it is impossible to predict future fluctuations of financial markets,
but in real financial markets, several statistical properties such as the weekend effect\cite{French}, the month-end effect\cite{Ariel},
and the holiday effect\cite{Lakonishok} have been discovered, which contradicts the efficient market hypothesis.
Therefore, these properties are called ``anomaly.''
Basically, the reason why anomalies occur cannot be explained,
but the Gotobi anomaly reported by Ref.\cite{akiyama} has a possible reason
to some extent.

Gotobi is a Japanese word and means the days divisible by five or ten,
that is, 
5, 10, 15, 20, 25, and 30 of each month.
As one of Japanese business customs,
there has been ``Goto-barai''
since before the Edo period,
and lots of Japanese companies still keep this business custom.
In particular, since domestic import companies mainly pay in US dollars,
the demand for selling the yen and buying the dollar by TTM (Telegraphic Transfer Middle Rate) that is the intraday exchange rate
tends to be increased at 9:55 a.m. when TTM is decided.
Moreover, since banks can know the demand in advance,
they try to purchase the dollars earlier and cheeper enough to deal with the demand, that is, the cover transaction,
which puts upward pressure on the USD/JPY rate toward 9:55. 
%
On the other hand,
since domestic export companies are waiting for payment from their counterparties
that do not have the business custom of Goto-barai,
the demand for buying the yen and selling the dollar does not occur in 
the Gotobi days.
Therefore, by
the asymmetry of demand, there might be the mechanism that
the yen becomes relatively cheaper than the dollar.
This phenomenon has been rumored in FX traders as an empirical rule of the USD/JPY market,
but Ref.\cite{akiyama} had confirmed it by statistical analyses with real data.

In the present study,
we discuss what kind of trading strategies FX traders including speculators come up with ideas on
after the anomaly became widely known.
The anomaly will disappear if Japanese companies stop the business custom of Goto-barai, 
the purpose of this research is to warn about the economic losses
caused by possible trading strategies performed by rational market participants
if the anomaly is left unattended.

In Section 2, we consider what kind of hypotheses might be derived
by FX traders who recognized the existence of the Gotobi anomaly
by analyzing the recent historical data of the USD/JPY rate during 2018 to 2020.
In Section 3, we consider the rational trading strategy that FX traders might come up with
by combining the hypotheses derived in Sect. 2 and some basic concepts of the technical analysis
to detect trading timings.
By confirming the effectiveness of this trading strategy,
we recognize that the Gotobi anomaly is a kind of arbitrage opportunity
and the possibility that the wealth of Japanese companies is being leaked to FX traders
owing to leaving the anomaly without attention.

For the analysis of our study, 
we use the order book information of the Electronic Broking System (EBS),
which enables us to analyze not only contract prices but also trading volumes in short time scale.
Since the previous study\cite{akiyama} analyzed in hourly time scale
by approximating 9:55 as 10:00,
the detailed analysis around 9:55 when TTM is decided has not yet been conducted.
%

\section{Statistical Properties Before and After 9:55}
We analyze what kind of hypotheses might be derived by FX traders
who recognized the existence of the Gotobi anomaly
by analyzing the recent historical data of the USD/JPY rate during 2018 to 2020
in order to make trading strategies.
For this analysis, we distinguish between Gotobi and non-Gotobi days,
and try to extract some statistical properties hidden in Gotobi days.
However, the number of Gotobi days is small, 
and therefore
the previous business days were included if Gotobi days were holiday  
as in our previous study\cite{akiyama}.
On the other hand, because the number of non-Gotobi days is large,
non-Gotobi days were randomly selected for the same number of Gotobi days.
Moreover, because the world's FX markets are closed early in the morning on Mondays in Japan,
we used daily data from Tuesday to Friday on business days for the sample of Gotobi and non-Gotobi days.


\begin{figure}[htbp]
\begin{center}
\includegraphics[width=0.9\linewidth,height=4.5cm]{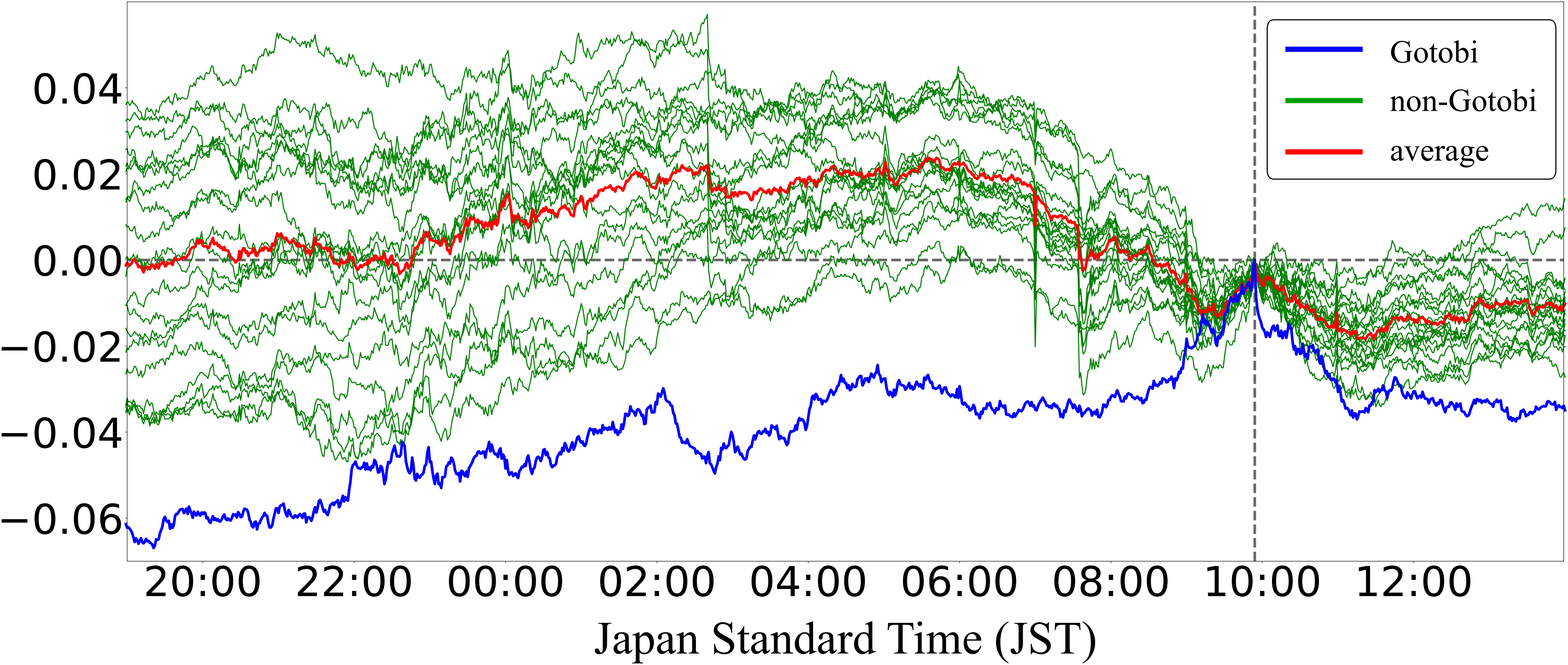} 
\caption{
Movement of the dollar-yen rate every minute during 1/1/2018 $\sim$ 12/31/2020:
the vertical axis at 9:55 (i.e., the decision and announcement time of TTM)
is set to 0.00.
The blue line is the average movement of Gotobi days, each green line is that of non-Gotobi days
randomly sampled for the same number of Gotobi days, and the red line is the average of all green lines. 
%
}
\label{figure:1}
\end{center}
\begin{center}
  \includegraphics[width=0.9\linewidth,height=5.5cm]{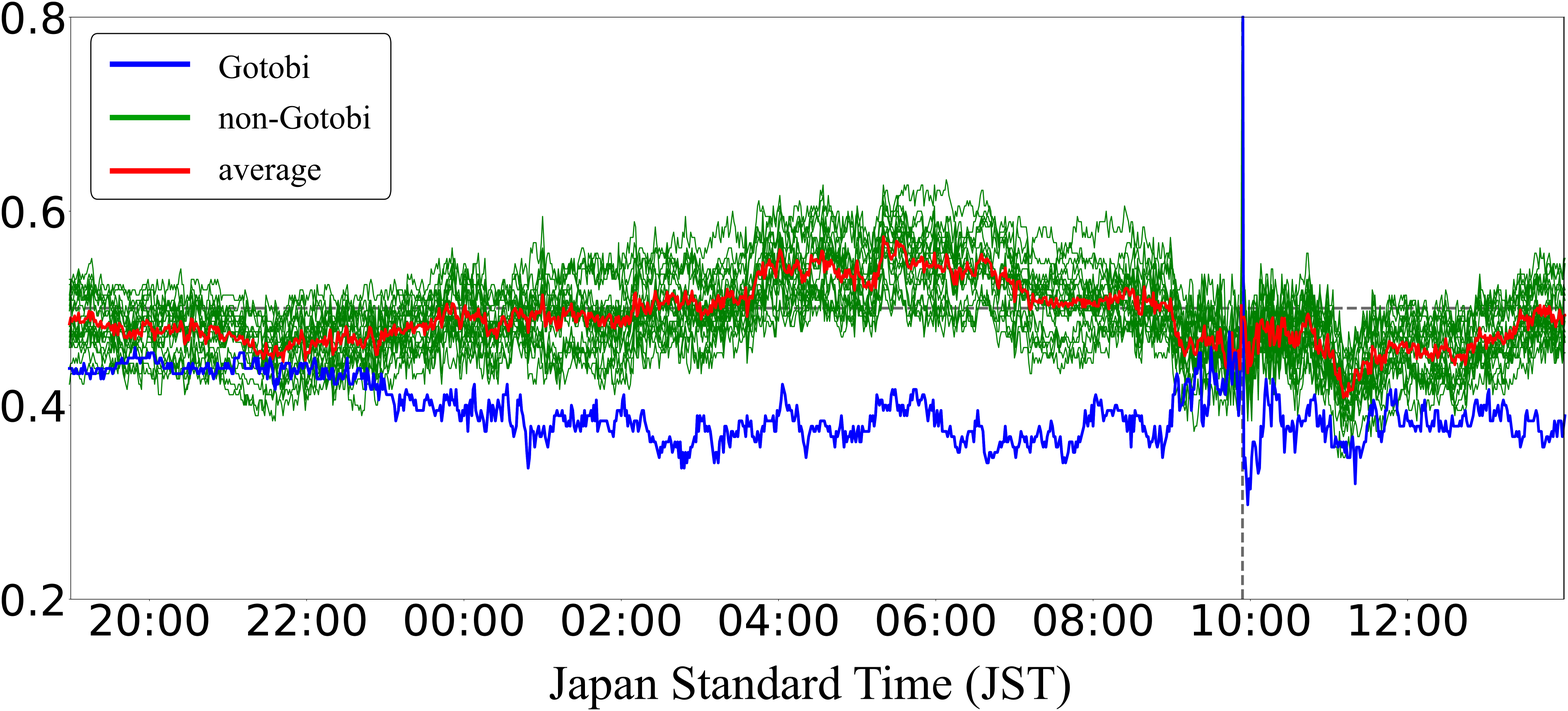}
\caption{
Probability that the dollar-yen rate every minute is more than that of 9:55 on the same day.
%
}
\label{figure:2}
\vspace{5mm}
\end{center}
\begin{center}
\includegraphics[width=0.9\linewidth,height=4.5cm]{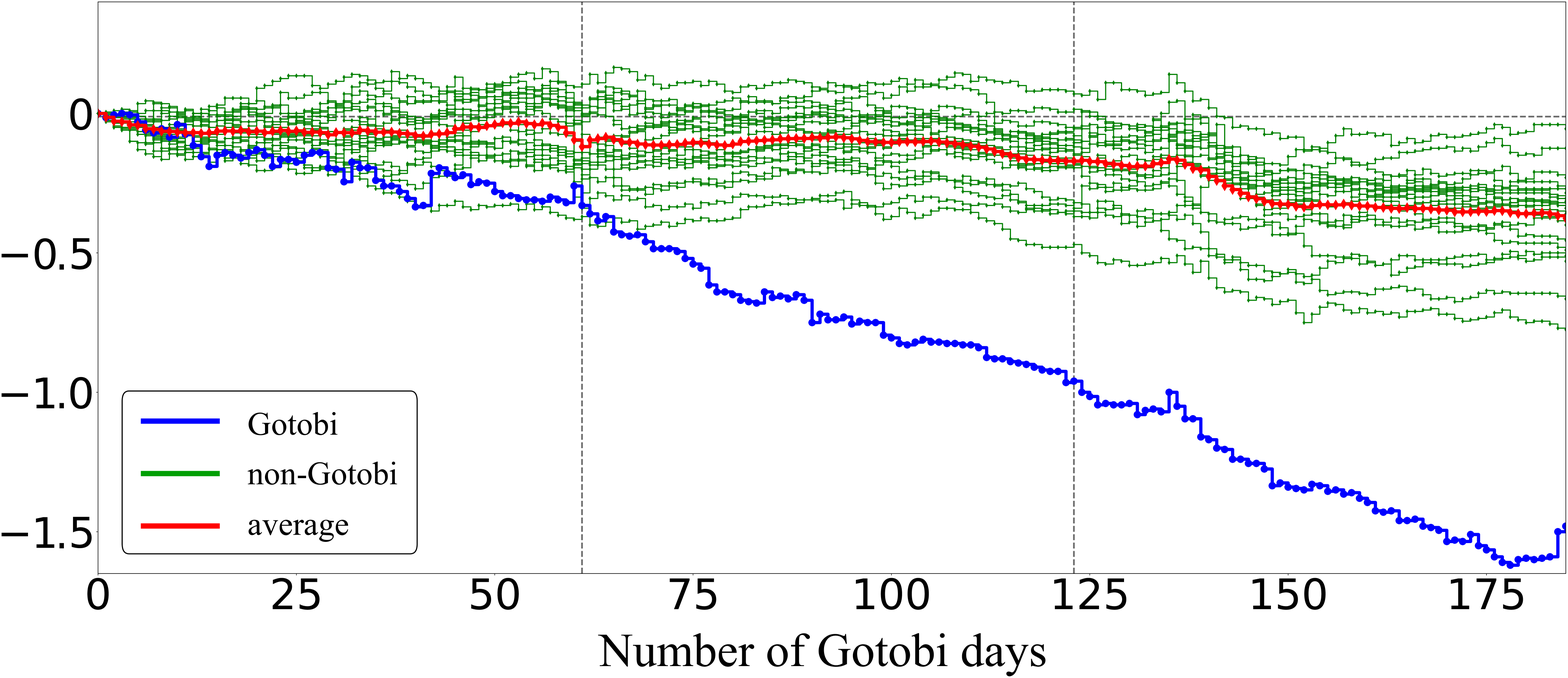}
\caption{
Cumulative movement of the dollar-yen rates during 9:55 to 9:56
(i.e., one minute just after the announcement of TTM at 9:55).
%
Two vertical dotted lines represent the fist days of 
2019 and 2020, respectively.
}
\label{figure:3}
\end{center}
\end{figure}

As shown in Figure \ref{figure:1},
the movement of the USD/JPY rate is clearly different between the Gotobi days and the non-Gotobi days.
In particular, 
we can confirm the Gotobi anomaly that the USD/JPY rate keeps increasing until 9:55 a.m. in the Gotobi days.
Next, Figure \ref{figure:2} shows the probability that the Gotobi anomaly gets started,
and the anomaly might be started around 3:00 a.m. (JST).
These results are consistent with the previous study\cite{akiyama},
and therefore we derive the first hypothesis as follows:
\begin{description}
\item[Hypothesis 1]
Entering the USD/JPY market by selling the yen and buying the dollar around 3 a.m. (JST) 
is likely to benefit from the Gotobi anomaly. 
\end{description}

Next, according to Figures \ref{figure:1} and \ref{figure:2},
the USD/JPY rate is suddenly decreased just after 9:55 a.m.
that is the decision and announcement time of TTM.
If the Gotobi anomaly is caused by the mispricing due to market overreaction,
the mispricing must be corrected by the market efficiency after 9:55
at which time the settlement demand of Japanese import companies is satisfied
and TTM is fixed.
As its evidence, 
Figure \ref{figure:3} shows the movement of the USD/JPY rates for a minute just after 9:55.
We can confirm the fact that the exchange rate is clearly decreased only in Gotobi days,
which means that 
the increase of exchange rates due to the Gotobi anomaly is caused by the mispricing,
and consequently its modification is stated just after 9:55.
From this viewpoint, we derive the second hypothesis as follows:
\begin{description}
\item[Hypothesis 2] The mispricing caused by the Gotobi anomaly is immediately corrected,
and therefore it is effective to reverse trading positions of buy and sell just after 9:55 when the occurrence of mispricing was comfirmed.
\end{description}



\section{Verification of Hypotheses by Technical Analysis}
We verify the validity of two hypotheses
and consider the rational trading strategy that ordinary FX traders might come up with,
assuming that FX traders use the moving average, which is the most popular tool of technical analysis,
to detect trading timings.
%
%
%
In order to calculate the profit obtained by trading strategy,
we subtracted the bid-ask spread when entering into the market and exiting from the market.
Then, the profit factor ($P_F$) was calculated by dividing the total of positive profits by the total of negative profits,
the payoff ratio ($P_R$) was calculated by dividing the average of positive profits by the average of negative profits,
and the winning percentage ($W$) was calculated by dividing the number of getting positive profits by the total number of all trades $N$.


\subsection{Verification of Hypothesis 1}
\begin{figure}[htbp]
\begin{center}
  \includegraphics[width=0.9\linewidth,height=4.5cm]{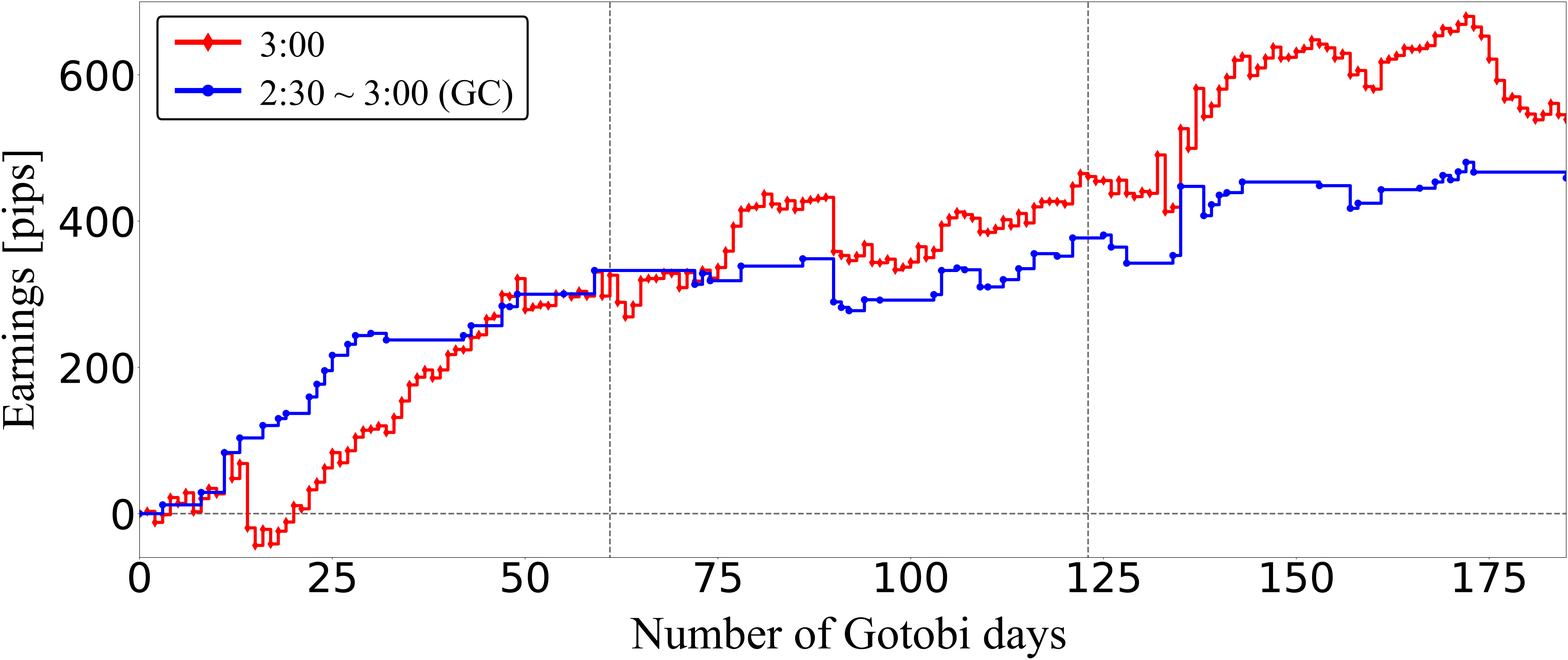}
\caption{
Cumulative earnings in Gotobi days by two cases of using the GC strategy during 2:30 to 3:00
and entering the market at 3:00 without the GC strategy.
%
In using the GC strategy,
$N=65$, $P_F = 2.62$, $P_R = 1.11$, and $W = 0.68$. 
In not using the GC strategy,
$N= 185$, $P_F = 1.46$, $P_R = 0.94$, and $W = 0.60$. 
}
 \label{figure:4}
\end{center}
\begin{center}
  \includegraphics[width=0.9\linewidth,height=4.5cm]{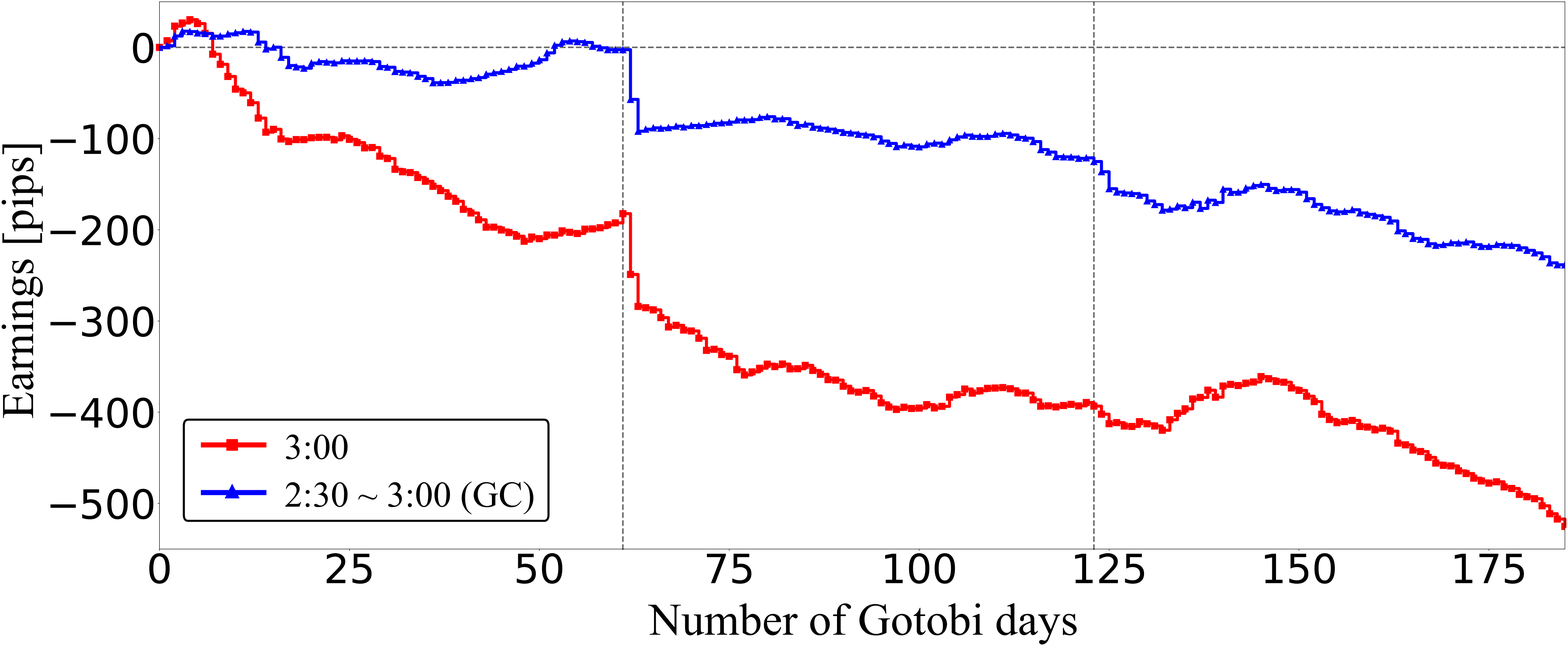}
\caption{
Same as Figure 4, but in non-Gotobi days.
In using the GC strategy, $N=69$, $P_F = 0.52$, $P_R = 0.60$, and $W = 0.46$.
In not using the GC strategy, $N=185$, 
$P_F = 0.51$, $P_R = 0.69$, and $W = 0.41$.
}
\label{figure:5}
\end{center}
\begin{center}
\includegraphics[width=0.9\linewidth,height=4.5cm]{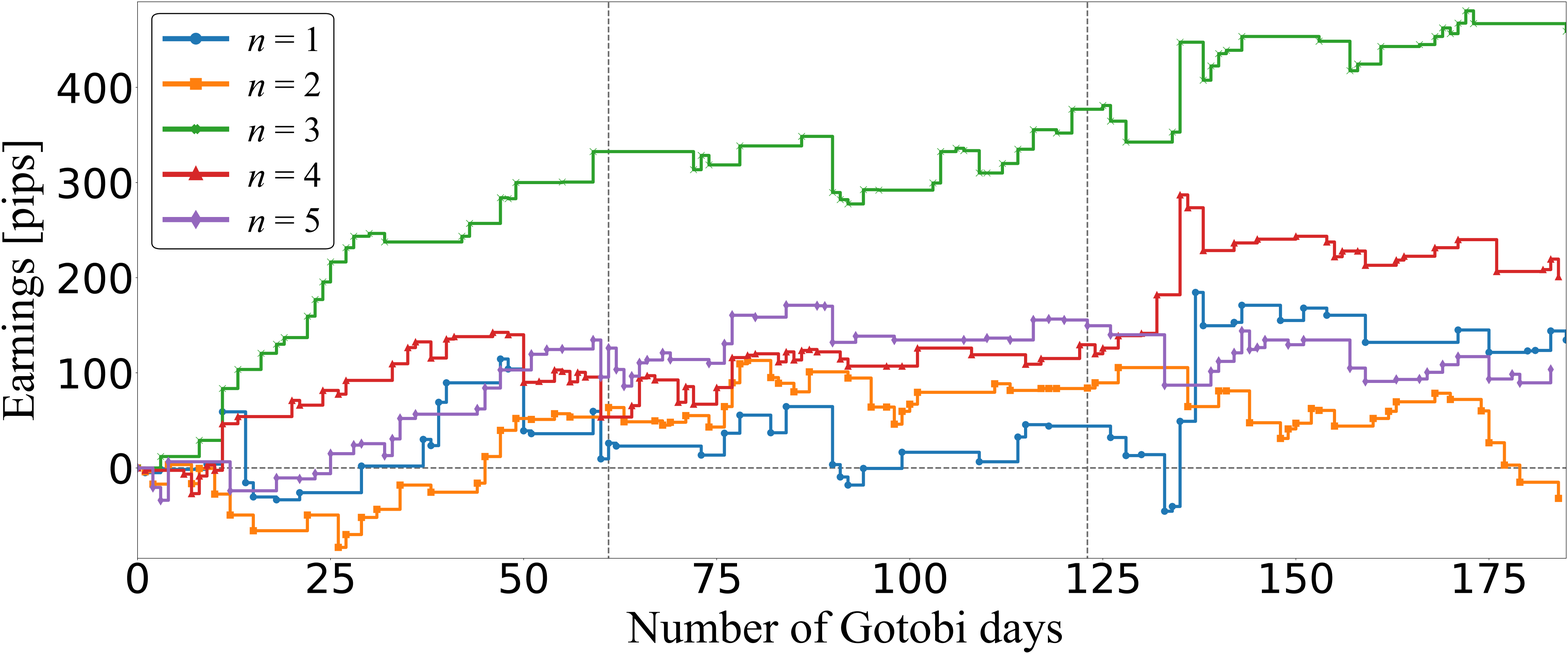}
\caption{
Same as Figure 4, but by the case of changing the time to apply the GC strategy during 30 minutes past $n-1$ to $n$.
If $n=1$, it is during 0:30 to 1:00.
When $n = 1$, $N=55$, $P_F = 1.24$, $P_R = 1.11$, and $W = 0.52$. 
When $n = 2$, $N=72$, $P_F = 0.94$, $P_R = 0.79$, and $W = 0.50$. 
When $n = 4$, $N=79$, $P_F = 1.47$, $P_R = 1.02$, and $W = 0.56$. 
When $n = 5$, $N=69$, $P_F = 1.27$, $P_R = 0.79$, and $W = 0.60$. 
}
 \label{figure:6}
\end{center}
\end{figure}

Figure \ref{figure:4} shows the profitability of entering the market at 3:00 and exiting it at 9:55 in the Gotobi days,
and its earnings are stably increasing, which can be an evidence of the first hypothesis.
On the other hand,
Figure \ref{figure:5} shows the same simulation but in non-Gotobi days,
and its earnings are stably decreased by the bid-ask spreads and are completely different in the Gotobi days.

If FX traders recognized the existence of the anomaly, 
they might try to optimize the timing of entering the market by using technical analysis as a time-series filter.
For this reason, we apply the golden cross (GC) based on 
the $25$-minute (short-term) moving average (SMA) and
the $100$-minute (long-term) moving average (LMA).
and enter the market if the GC occurs during 2:30 to 3:00 and then exit it at 9:55. 
As shown in Figures \ref{figure:4} and \ref{figure:5},
using the GC can improve the total earnings and trading efficiency of the Gotobi days
even though the number of trades $N$ is reduced.

Figure \ref{figure:6} shows the case of changing the time to apply the above GC strategy.
As a result, the performance of $n = 3$ is the best, which also shows the validity of the first Hypothesis.

\subsection{Verification of Hypothesis 2}
If the Gotobi anomaly, that is, the irregular increase of exchange rate is caused by the mispricing of market,
rational traders might attempt to reverse their buy positions into sell positions for following its modification process.
Therefore, we verify the profitability of taking a sell position just after 9:55 and closing its position at 12:00.
Here, because the occurrence of the Gotobi anomaly is a necessary condition for the mispricing,
this strategy is performed only when the profit obtained until 9:00 is more than $0$.

\begin{figure}[htbp]
\begin{center}
  \includegraphics[width=0.9\linewidth,height=4.5cm]{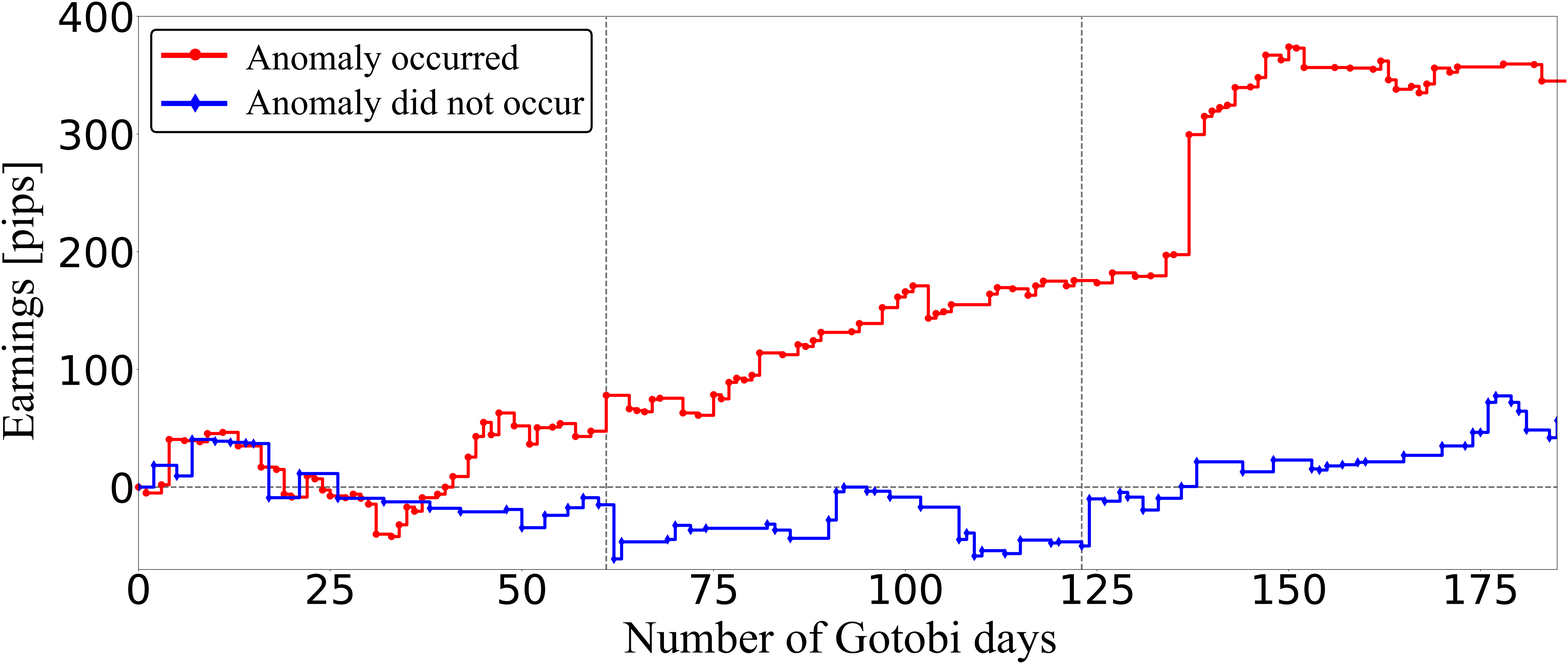}
\caption{
Cumulative earnings in Gotobi days by taking a sell position
during 9:55 to 12:00.
In the days when the Gotobi anomaly occurred, 
$N = 113$, $P_F = 2.09$, $P_R = 1.51$, and $W = 0.57$.
In the days when the Gotobi anomaly did not occur, 
$N = 72$, $P_F = 1.18$, $P_R = 1.15$, and $W = 0.48$.
}
\label{figure:7}
\end{center}
\begin{center}
  \includegraphics[width=0.9\linewidth,height=4.5cm]{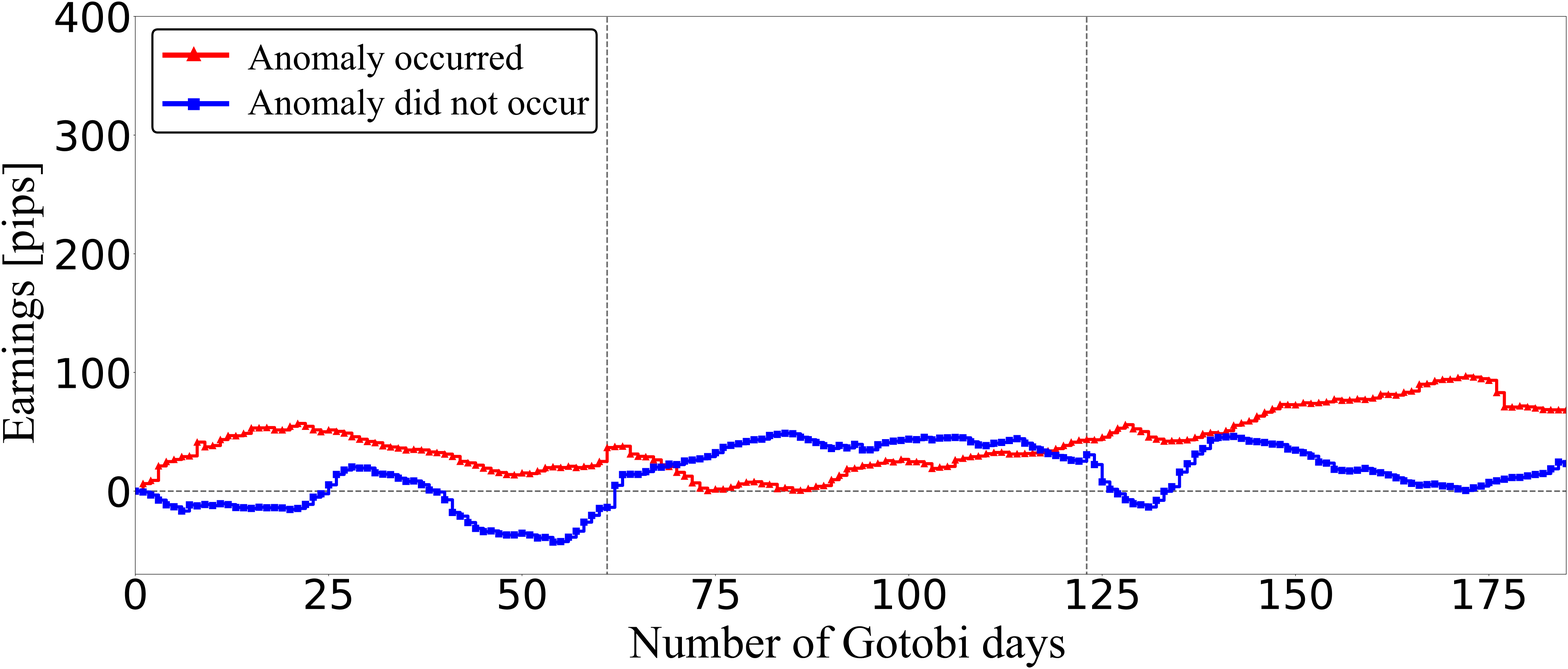}
\caption{
Same as Figure 7, but in non-Gotobi days.
In the days when the Gotobi anomaly occurred, 
$N = 88$, $P_F = 1.20$, $P_R = 1.07$, and $W = 0.51$.
In the days when the Gotobi anomaly did not occur, 
$N = 72$, $P_F = 1.08$, $P_R = 1.11$, and $W = 0.47$.
  }
 \label{figure:8}
\end{center}
\begin{center}
  \includegraphics[width=0.9\linewidth,height=4.5cm]{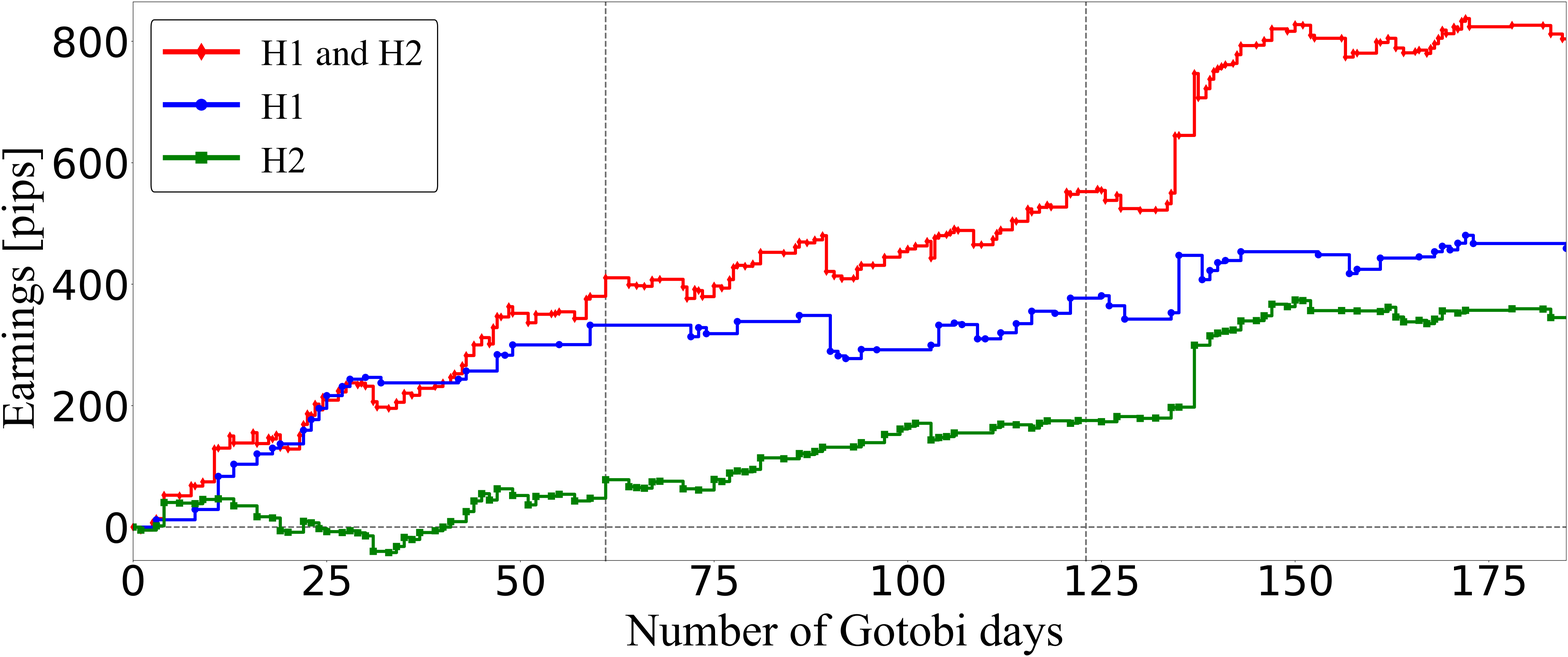}
  \caption{
Cumulative earnings in Gotobi days by the combination of two hypotheses: H1 and H2
where $N = 167$,  $P_F = 2.60$, $P_R = 1.48$, and $W = 0.63$.
Here, H1 corresponds to the blue-colored line in Figure 4,
and H2 corresponds to the red-colored line in Figure 7.
}
 \label{figure:9}
\end{center}
\end{figure}

Figure \ref{figure:7} shows the profits of taking a sell position during 9:55 to 12:00.
If the Gotobi anomaly occurs, this strategy works well.
However, even in Gotobi days, this strategy is meaningless if the Gotobi anomaly does not occur.
Therefore, the usefulness of taking sell positions is caused by the Gotobi anomaly,
which also concludes that the anomaly is caused by the mispricing of market
and supports the second hypothesis.
Figure \ref{figure:8} shows the same simulation but in non-Gotobi days.
Even if the profit obtained until 9:00 is more than $0$, which is shown as ``Anomaly occurred'',
this strategy does not work because it has nothing to do with the mispricing caused by the Gotobi anomaly.

\subsection{Combination of Two Hypotheses}
Finally, by combining the trading strategies based on two hypotheses,
Figure \ref{figure:9} shows the final earnings that ordinary but rational FX traders can obtain.
As a result, the combination can get the most stable and largest profits by making the most of the Gotobi anomaly,
which means the possibility that the wealth of Japanese companies leaks to FX traders
if they blindly keep making payments in the Gotobi days as a business custom.
 
\section{Conclusion}
In this study, we presented two basic hypotheses that can be derived by FX traders who recognized the Gotobi anomaly,
and considered trading strategies that combines the hypotheses with the golden cross of popular technical analysis.
By confirming the usefulness of trading strategies through investment simulations,
it can be concluded that the Gotobi anomaly corresponds to a kind of arbitrage opportunity.
This is caused by the traditional business custom of Japanese import companies 
that make payments by TTM in the Gotobi days,
and therefore maintaining this business custom continues to drain their financial wealth to ordinary FX traders.
The scale of its loss is not necessarily huge, but it should be noted that it leaks stably.

This research was partially supported by a JSPS Grant-in-Aid for Scientific Research (20K11969). 
The contents of this article are the personal views of its authors and not the official views of the institutions with which they are affiliated.

\end{document}